\documentclass[12pt, preprint,dvips]{aastex}
\usepackage[]{graphicx}
\newcommand{\gray}{$\gamma$-ray }
\newcommand{\grays}{$\gamma$-rays }
\begin{document}
\title{Multiwavelength Observations of the Blazar Mrk~421 in December 2002 and January 2003 }
\author{ P.~F.~Rebillot,\altaffilmark{1} 
	      H. M. Badran,\altaffilmark{2}
	      G. Blaylock,\altaffilmark{3}
	      S. M. Bradbury,\altaffilmark{4}
	      J. H. Buckley,\altaffilmark{1}
	      D. A. Carter-Lewis,\altaffilmark{5}
	      O. Celik,\altaffilmark{6}
              Y.C. Chow,\altaffilmark{6}
              P. Cogan,\altaffilmark{7}
	      W. Cui,\altaffilmark{8}
	      M. Daniel,\altaffilmark{7}
	      C. Duke,\altaffilmark{9}
	      A. Falcone,\altaffilmark{8}
	      S. J. Fegan,\altaffilmark{6}
	      J. P. Finley,\altaffilmark{8}
	      L. F. Fortson,\altaffilmark{10,11}
	      G. H. Gillanders,\altaffilmark{7}
	      K. Gutierrez,\altaffilmark{1}
	      G. Gyuk, \altaffilmark{10,11}
	      D. Hanna,\altaffilmark{12}
	      J. Holder,\altaffilmark{4}
	      D. Horan,\altaffilmark{13}
	      S. B. Hughes, \altaffilmark{1}
	      G. E. Kenny,\altaffilmark{7}
	      M. Kertzman,\altaffilmark{14}
	      D. Kieda,\altaffilmark{15}
	      J. Kildea,\altaffilmark{12}
	      K. Kosack,\altaffilmark{1}
	      H. Krawczynski,\altaffilmark{1}
	      F. Krennrich,\altaffilmark{5}
	      M. J. Lang,\altaffilmark{7}
	      S. Le Bohec,\altaffilmark{15}
	      E. Linton,\altaffilmark{16}
	      G. Maier,\altaffilmark{4}
	      P. Moriarty,\altaffilmark{17}
	      J. Perkins,\altaffilmark{1}
	      M. Pohl, \altaffilmark{5}
	      J. Quinn,\altaffilmark{7}
	      K. Ragan,\altaffilmark{12}
	      P. T. Reynolds,\altaffilmark{18}
	      H. J. Rose,\altaffilmark{4}
	      M. Schroedter,\altaffilmark{5,13}
	      G. H. Sembroski,\altaffilmark{8}
	      G. Steele,\altaffilmark{11}
	      S. P. Swordy,\altaffilmark{16}
	      L. Valcarcel,\altaffilmark{12}
	      V. V. Vassiliev,\altaffilmark{6}
	      S. P. Wakely,\altaffilmark{16}
	      T. C. Weekes,\altaffilmark{13}
	      J. Zweerink \altaffilmark{6} (The VERITAS Collaboration)}
\and
\author{M. Aller \altaffilmark{19}, H. Aller \altaffilmark{19},
P. Boltwood \altaffilmark{20},
I. Jung \altaffilmark{1},
D. Kranich \altaffilmark{21},
A. Sillanpaa  \altaffilmark{22},
A. Sadun  \altaffilmark{23}}
              \altaffiltext{1}{Department of Physics, Washington University, 
		St. Louis, MO 63130, USA}  
	      \altaffiltext{2}{Physics Department, Tanta University,
		Tanta, Egypt}
	      \altaffiltext{3}{Department of Physics, University of
		Massachussetts, Amherst, MA 01003-4525, USA}
	      \altaffiltext{4}{School of Physics and Astronomy, University of
		Leeds, Leeds, LS2 9JT, Yorkshire, England, UK}
	      \altaffiltext{5}{Department of Physics and Astronomy,
		Iowa State University, Ames, IA 50011-3160, USA}
	      \altaffiltext{6}{Department of Physics, University of
		California, Los Angeles, CA 90095-1562, USA}
	      \altaffiltext{7}{Department of Physics, National
		University of Ireland, Galway, Ireland}
	      \altaffiltext{8}{Department of Physics, Purdue
		University, West Lafayette, IN 47907, USA}
	      \altaffiltext{9}{Department of Physics, Grinnell
		College, Grinnell, IA 50112-1690, USA}
	      \altaffiltext{10}{Department of Astronomy and Astrophysics,
		University of Chicago, Chicago, IL, USA}
	      \altaffiltext{11}{Astronomy Department, Adler Planetarium and
		Astronomy Museum, Chicago, Il, USA.}
	      \altaffiltext{12}{Physics Department, McGill University, 
		Montreal, QC\,H3A\,2T8, Canada}
	      \altaffiltext{13}{Fred Lawrence Whipple Observatory,
		Harvard-Smithsonian CfA, P.O. Box 97, Amado, AZ 85645-0097} 
	      \altaffiltext{14}{Department of Physics and Astronomy,
		DePauw University, Greencastle, IN 46135-0037, USA}
	      \altaffiltext{15}{High Energy Astrophysics Institute,
		University of Utah, Salt Lake City, UT 84112, USA}
	      \altaffiltext{16}{Enrico Fermi Institute, University of
		Chicago, Chicago, IL 60637, USA}
	      \altaffiltext{17}{School of Science, Galway-Mayo
		Institute of Technology, Galway, Ireland}
	      \altaffiltext{18}{Department of Applied Physics and
		Instrumentation, Cork Institute of Technology, Cork, Ireland}
	      \altaffiltext{19}{University of Michigan, Ann Arbor, MI 48109, 
		USA}
	      \altaffiltext{20}{Boltwood Observatory, Ontario, Canada}
	      \altaffiltext{21}{University of California, Davis, 
		CA 95616, USA}
	      \altaffiltext{22}{Institute for Space Physics, 
		University of Turku, Vaisala, Finland}
	      \altaffiltext{23}{Department of Physics, 
		University of Colorado at Denver and Health Sciences Center,
		Denver, CO 80208, USA}
\email{\small corresponding authors: Paul Rebillot <rebillot@physics.wustl.edu> 
and Henric Krawczynski <krawcz@wuphys.wustl.edu>}
\begin{abstract}
We report on a multiwavelength campaign on the TeV \gray blazar 
Markarian (Mrk) 421 performed during December 2002 and January 2003.
These target of opportunity observations were initiated by the 
detection of X-ray and TeV \gray flares with the All Sky Monitor 
(ASM) on board the {\it Rossi X-ray Timing Explorer} 
({\it RXTE}) and the 10~m Whipple \gray telescope.
The campaign included observational coverage in the radio 
(University of Michigan Radio Astronomy Observatory),
optical (Boltwood, La Palma KVA 0.6m, WIYN 0.9m), X-ray (RXTE pointed telescopes), 
and TeV \gray (Whipple and HEGRA) bands.  

At TeV energies, the observations revealed several flares at 
intermediate flux levels, peaking between 1 and 1.5 times the 
flux from the Crab Nebula. While the time averaged spectrum can be 
fitted with a single power law of photon index $\Gamma =2.8$
from $dN_\gamma/dE\propto E^{-\Gamma}$, we find some evidence 
for spectral variability.
Confirming earlier results, the campaign reveals a rather loose 
correlation between the X-ray and TeV \gray fluxes.
In one case, a very strong X-ray flare is not accompanied by
a comparable TeV \gray flare.
Although the source flux was variable in the optical and radio 
bands, the sparse sampling of the optical and radio light 
curves does not allow us to study the correlation properties 
in detail. 

We present a simple analysis of the data with a 
synchrotron-self Compton model, emphasizing that models with very high
Doppler factors and low magnetic fields can describe the data.
\end{abstract}

\keywords{BL Lacertae objects: individual (Mrk421),
  galaxies: jets, gamma rays: observations, radiation mechanisms:
  non-thermal, X-rays: individual (Mrk421) }
\section{Introduction}
\paragraph{}
The space-borne EGRET {\it (Energetic Gamma Ray Experiment Telescope)} 
detector on board the {\it Compton Gamma-Ray Observatory} 
discovered strong MeV and GeV \gray emission from 66 blazars, 
mainly from Flat Spectrum Radio Quasars and Unidentified 
Flat Spectrum Radio Sources  \citep{Hart:92,Hart:99}.
Ground-based Cherenkov telescopes discovered TeV \gray 
emission from seven blazars, five of which were not detected by 
EGRET \citep{Week:04,Ahar:05}. 
Although \gray emission from blazars have been studied for more 
than a decade now, it is still unclear where and how the emission originates.
According to the most common paradigm, the emission originates close
to a mass-accreting supermassive black hole, in a relativistically
moving collimated plasma outflow (jet) that is aligned with the line
of sight to within a few degrees. The relativistic Doppler effect can 
explain the intensity of the blazar emission, and its rapid variability at 
X-rays and \gray energies on hour time scales:
for emission originating from synchrotron or synchrotron self-Compton
(SSC) models,
the apparent luminosity increases approximately as the fourth power of the 
relativistic Doppler factor\footnote{The relativistic Doppler factor is given by 
$\delta_{\rm j}\,=$ $\left[\Gamma(1-\beta\,\cos{\theta})\right]^{-1}$ 
with $\Gamma$, the bulk Lorentz factor of the jet plasma, $\theta$, the angle 
between jet axis and the line of sight, and $\beta$, the plasma
velocity, in units of the speed of light.}
$\delta_{\rm j}$, and the observed flux variability timescale is 
inversely proportional to $\delta_{\rm j}$.

Blazars are powerful sources across the electromagnetic spectrum.
Typical spectral energy distributions (SEDs) for the high energy
peaked TeV blazars show two broad peaks,
one at infrared to X-ray energies and the other at X-ray to \gray energies.
The low-energy peak is commonly believed to originate as synchrotron 
emission from a population of relativistic electrons gyrating 
in the magnetic field of the jet plasma. The origin of the 
high-energy peak is unproven. The commonly adopted and 
best studied models assume that the \grays are 
produced in inverse Compton processes by the same electrons 
that emit the synchrotron radiation at longer wavelengths  
(for a recent review of observations and models, see \citet{Kraw:04r}).
In so-called hadronic models, \grays are emitted as synchrotron 
radiation of extremely energetic protons \citep{aharonian,muecke}, 
as inverse Compton and synchrotron emission from a Proton Induced Cascade 
(PIC) \citep{manheim}, or from $\pi^{0}\rightarrow$ $\gamma\gamma$ 
decays following the interaction of high energy protons with 
some target material \citep{Pohl:00}. 
Recent reviews on observations of blazars with TeV emission 
and models developed to describe the data can be found in various 
review articles and books \citep{Kraw:04r,Kraw:05,Tave:04,Trev:03,Feli:04}.
Reviews focussing on observations and models of sources with  
MeV/GeV emission can be found in \citet{Siko:01b,Copp:99}. 
Broader overviews of the field of TeV \gray astronomy 
are given in \citet{Buck:02,Ong:03,Trev:03,Feli:04}.

Multiwavelength observations are key for understanding the blazar
phenomenon. The acquisition of good multiwavelength data sets has 
encountered substantial difficulties as the sensitivities of current 
TeV observatories require flares for sampling the TeV light curves 
on a time scale of hours. 
Some sources were observed with excellent multiwavelength coverage but 
during relatively unspectacular quiescent phases; in other cases, the 
sources were flaring, but the fluxes were only poorly sampled in 
frequency space and in time. The most remarkable result from the 
multiwavelength campaigns is that there is good evidence for a 
correlation between the X-ray fluxes and the TeV \gray fluxes 
for the two sources Mrk 421 \citep{Buck:96,Taka:96,Taka:00,Blaz:05} and 
Mrk 501 \citep{Djan:99,Samb:00,Kraw:02}.

In this paper we present results from a multiwavelength campaign on the
TeV blazar Markarian (Mrk) 421. The source is a nearby (z = 0.031) 
high energy-peaked BL Lac object, and was the first extragalactic
source detected in the TeV \gray band \citep{Punc:92}. 
In November 2002, \gray observations with the Whipple 10~m telescope 
revealed several Mrk 421 flares with fluxes exceeding three times 
the steady flux from the Crab Nebula. The All Sky Monitor 
instrument aboard {\it RXTE} also showed extremely strong 
2-12 keV fluxes reaching 100 milli-Crab. Collectively these triggered
a coordinated campaign. We invoked radio, optical, and X-Ray ({\it RXTE})
observations to commence as soon as the waning Moon would allow 
the Cherenkov telescope to take data once more.
Although the X-ray and TeV \gray fluxes had decreased 
substantially when the campaign started on December 4th, 
we acquired a data set with a high signal-to-noise-ratio 
X-ray light curve and X-ray energy spectra, and good 
signal-to-noise-ratio TeV light curves and TeV \gray
energy spectra. The combined data allowed us to study the 
X-ray/TeV \gray flux correlation over several weeks. 
Following our previous study \citep{Blaz:05}, this
is the second campaign that measures the X-ray/TeV \gray flux
correlation over several weekes. For the first time, 
we use here simulated lightcurves to address the statistical 
significance of the X-ray/TeV \gray flux correlation and 
to constrain the time lag between the two light curves.
Simulations are necessary as subsequent data points in
the light curve are not independent of each other 
(see e.g., the discussion by \citet{edelson}). The rest of the paper is organized
as follows.
After describing the data sets in Section 2, we explain the 
method that we used to reconstruct TeV \gray energy spectra
from the Whipple data, and give the results obtained with the 
method in Section 3. Subsequently, we present the results of the 
campaign in Section 4, and conclude with a summary and a discussion in 
Section 5.  If not mentioned otherwise, errors are quoted on the level 
of one standard deviation, and upper limits are given on 
90\% confidence level.
\section{Observations and Standard Data Reduction}
\subsection*{Radio Observations}
\paragraph{}

We used the University of Michigan equatorially mounted 26-meter
paraboloid in its automatic observation mode to observe Mrk 421 at 4.8
GHz, 8.0 GHz and 14.5 GHz between December 3, 2002 and January 10,
2003.  Both linear polarization and total flux density measurements
were made, but only the total flux density measurements are reported
here.  Each observation consisted of a series of five-minute ON-OFF
type measurements over a 40-minute time period interleaved with
observations of other program sources. Observations were restricted to
within three hours of the local meridian to minimize instrumental
errors.  The telescope pointing corrections are interpolated from
position scans through sources stronger than 2-Janskys.   The flux
density measurements have been corrected to the Baars flux density
calibration scale \citep{baars} using observations of a grid of
calibrator sources, distributed around the sky, which were observed
at approximately two-hour intervals. The $1\sigma$ error bars include
both the estimated measurement and calibration uncertainties. The
observation and calibration procedures have been described in more
detail elsewhere \citep{aller}.
 
\subsection*{Optical Observations}
\paragraph{}
In the following, we will discuss three optical data sets.
The first was taken at the Boltwood Observatory (Stittsville, 
Ontario, Canada) with a 0.4~m telescope, a SiTe 502A CCD chip, 
and a Johnson-Cousins R filter. The data were collected 
for $\sim$ 10 days between December 3, 2002 and January 12, 2003.  
Relative aperture photometry was performed with an aperture of 10 arcsec
and ``star 1'' from \citet{villata} as the comparison star. 
The background was estimated using a concentric annulus 
with a diameter between 37 and 44 arcsec. We did not subtract 
the light from the host-galaxy.
Usually, five two-minute exposures were integrated before 
deriving the photometric value. The typical statistical 
error on the relative photometry is 0.02 mag.  The analysis 
is compromised by two very bright stars near Mrk 421 that cause a 
varying level of light to spill into the source and background regions. 
Based on previous optical results on the same source and stars 
in the field of view, we estimate that photometric measurements
have a systematic error of 0.08 mag per data point.
\paragraph{}
The second set of optical observations were made using a 35~cm 
Celestron telescope installed on the tube of the 60~cm KVA telescope
(La Palma, Spain). The observations were made with the ST-8 CCD using 
a standard Kron/Cousins R-filter. The analysis used the reduction 
programs developed by Kari Nilsson (Tuorla Observatory) with
the reference stars one and two from \citet{villata}.  
\paragraph{}
The final set of optical data were collected using the WIYN 0.9 m
telescope at KPNO with the S2KB CCD imager using a Harris V-filter.   
The data were collected from 6 December 2002 to 15 January 2003 
(with, however, a large gap 9 Dec 2002 to 3 Jan 2003).  
Relative aperature photometry was performed using standard IRAF routines
with an aperature of 6 arcsec, sky annulus ranging from 27 to 30 arcsec in
diameter, and ``star 1'' of \citet{villata} as the comparison star. 
Again, we did not subtract light from the host galaxy.  Typical statistical
errors from the photometric fits were smaller than 0.005 mag.  Based on
the spread in magnitude difference between two reference stars, we
estimate the uncertainty for each data point to be 0.02 mag for the
purpose of determining variability.  With regard to the absolute flux, due
to the presence of host-galaxy light, we expect the values reported to
contain an undetermined systematic offset of as much as 0.1 mag.
Optical magnitudes for all three data sets are 
converted to absolute fluxes according to \citet{allen}.

\subsection*{X-ray Observations}
\paragraph{}
We reduced the data from the {\it RXTE} Proportional Counter Array
with the standard {\it RXTE} data analysis software.
Standard-2 mode PCA data taken with the top layer of the operational  
Proportional Counter Units (PCUs) were analyzed. The number of PCUs 
operational during a pointing varied between 2 and 4.
We restricted the spectral analysis to the energy range from 4 keV to 15 keV.
We excluded data below 4 keV, as the analysis of earlier {\it RXTE} data
showed corrupted behavior (exceptionally high or low count rates of 
individual bins not compatible with the energy resolution of the instrument). 
We find that the data of most pointings are dominated by background 
above 15 keV. After applying the standard screening 
criteria (including visual inspection of the electron rate) and removing 
abnormal data spikes, the net exposure in each Good Time Interval 
ranged from 168 sec to 9.01 ksec. Light-curves were then extracted
with \verb+FTOOLS+ v5.3.2. background models were generated with the tool 
\verb+pcabackest+ , based on the {\it{\it RXTE}} GOF calibration files 
for a ``bright'' source (more than 40 counts/sec/PCU). Response
matrices for the PCA data were created with the script 
\verb+pcarsp+ v.10.1. 
The PCU ``PCU0'' was not excluded for analysis as the FTOOLS version 
gives the proper background model. 
We assume for all fits a galactic neutral hydrogen density of 
1.31 $\times10^{20}$ cm$^{-2}$ 
\citep{nrao}\footnote{http://asc.harvard.edu/toolkit/colden.jsp}.
For each pointing, a power law model was fitted over the energy range 
from 4~keV to 15~keV.

\paragraph{}
We complement the data from the pointed RXTE telescopes with data from the
RXTE All Sky Monitor (ASM) \citep{levine} taken between December 2, 2002 (UT)
and January 14, 2003 (UT). We derived fluxes by averaging the 
``summed band intensities'' acquired during one day. 

\subsection*{TeV \gray Observations}\label{gray_analysis}
TeV observations were taken with the Whipple 10 m Cherenkov
telescope (Mount Hopkins, AZ) and with the CT1 telescope of the
High Energy Gamma-Ray Astronomy (HEGRA) collaboration (La Palma, Spain). 
In the following paragraphs we describe the two data sets.

The Whipple observations were taken between 4 December 2002 (UT) 
and 15 January 2003 (UT). A total of 44~hrs of data were acquired:
32 hrs on the source, and 12 hrs on an adjacent field for background
estimation purposes. The data were analyzed using the standard ``Hillas'' 
2$^{\rm nd}$-moment-parameterization technique \citep{Hill:85}. 
Standard cuts ($\verb SuperCuts 2000$) were used to 
select \gray events and to suppress background cosmic-ray events
\citep{Call:03}.
The fluxes were normalized to the flux from the Crab Nebula 
using a data set of 15 hrs of on-source data and matching background 
observations taken in December 2002 and January 2003
\citep{Punc:91}. 
Using the zenith angle dependence of this Crab data set we account for
the zenith angle dependence of the \gray excess rate  by
normalizing our measured Mrk421  rate to the Crab rate at a
corresponding zenith angle.
Significances and corresponding error bars
were calculated using the method of \citet{lima}.

From Monte Carlo simulations, we fold the Crab spectrum with the
instrument response to obtain the peak energy (energy threshold) for
the Whipple 10m data to be consistent with the value 400 GeV derived
elsewhere \citep{Petr:02}.
More detailed descriptions of Whipple observing modes and analysis 
procedures can be found elsewhere \citep{weekes,punch,reynolds}.
Details about the Whipple telescope including the GRANITE-III camera
have been given in \citet{finley}.  

\paragraph{}
A second TeV \gray data set was acquired with the 
HEGRA CT1 telescpe (see \citet{rauterberg} for a description of the
CT1 instrument)
between 3 November 2002 and 12 December 2002. 
The telescope was equipped with a 127 pixel camera with a 3 degree 
diameter field of view, and with all-aluminum mirrors giving a 
total of $10 \, \mathrm{m^2}$ reflecting surface \citep{mirzoyan}. 
The analysis used 17 hrs of data with zenith angles between 12$^\circ$ 
and 58$^\circ$. 
The HEGRA CT1 data were normalized to the Crab flux in a similar 
way as the Whipple data, taking into account the zenith angle 
dependence of the excess rate. We estimate a mean energy threshold 
for the CT1 data set of approximately 700 GeV.

\paragraph{}
The normalization to the steady Crab flux is a convenient way to 
combine data from different instruments to avoid systematic 
errors resulting from errors in the absolute flux calibration 
of each instrument, and to perform a first-order correction for
variations in rate with zenith angle. The drawback of the method is that the 
different energy thresholds of the Whipple and HEGRA observations 
can introduce a normalization error if the source energy spectrum 
deviates from the Crab energy spectrum. 
Using previous measurements of the Mrk 421 TeV spectral index as a
 function of flux level, we estimate that the Whipple/HEGRA 
normalization error is always smaller than 30\%. 
\section{Determination of TeV \gray Energy Spectra with the Forward Folding Method}
The spectral analysis of the Whipple TeV \gray data used a different set of
gamma-hadron separation cuts
that minimize the systematic error associated with uncertainties in the
\gray selection efficiency of the cuts while still giving a good
sensitivity. The ``extended zenith angle scaled cuts'' \citep{kosack04} 
select primary \grays with an efficiency that is largely independent 
of the zenith angle of the observation and the energy of the primary photon.
The analysis is based on the Grinnell-ISU (GrISU) 
package\footnote{ http://www.physics.utah.edu/gammaray/GrISU}
that uses the KASCADE airshower simulation code \citep{Kert:94}, 
followed by the simulation of the Cherenkov light emitted 
by the air shower and the simulation of the detector response.
To calibrate the overall gain of the Whipple 10m telescope in the simulations, 
we compared simulated and observed muon events. Muons show up as bright 
arcs of Cherenkov light in the camera and are useful for calibration 
because the light per unit arc length is nearly constant, 
regardless of the impact parameter and angle of the muon trajectory. 
The overall gain of the telescope can be found by comparing the 
distribution of the signal per arc length in a simulated set of muon 
events and in an observed set. We took the simulated muon events from 
a sample of simulated proton and helium showers. 
Muons are identified with a dedicated muon identification algorithm that 
extracts $\sim$200 muon arcs per 28 min data run. 
We adjusted the overall gain factor in the simulations until they 
reproduced the observed  signal per arc length distribution. 
The overall gain factor agrees to within 15\% with the value computed 
from measurements of the mirror reflectivity, photo-multiplier tube
(PMT) quantum efficiency, and electronic gain.

We used the forward folding technique to fit the energy spectra.
Although earlier TeV \gray analyses used similar methods, 
this is the first time that we describe the method in detail.
For each Cherenkov event that passed the gamma-hadron separation cuts,
we computed an estimator $E$ of its primary energy, based on 
the image parameters `{\it size}' $S$ (sum of counts in an image) and `{\it distance}' 
$d$ (distance of the image centroid from the center of the 
field of view):
\begin{equation}
\ln{E}\,\,=\,\,g(x)+h(d) \,\,\, ,
\end{equation}
with $x\,=\ln{S}$, $g(x)\,=$ $A\,+\,B\,x\,+\,C\,x^2$ and
$h(d)\,= D_1\,+\,\alpha\,d$ for $d<d_0$ and 
$h(d)\,= D_2\,+\,\beta\,d$ for $d>d_0$.
The constants $A$, $B$, $C$, $D_1$, $D_2$, $\alpha$, $\beta$, and $d_0$ are 
given in Table \ref{parameters}. 
The first term in the energy estimator reflects the fact that
total intensity of an image ({\it size}) is roughly proportional
to the energy ($E$) of the inducing \gray. The second term
corrects this relationship depending on the distance of the
telescope from the shower axis ($d$ is proportional to the latter).
Using extended zenith angle cuts, the
energy estimator gives an energy resolution of $\sigma(\ln{E})\,=\,$ 0.25.

We limited the spectral analysis to ON-source data with associated background data sets
(the so-called ON-OFF data) taken at zenith angles less than 30$^\circ$. 
After histogramming the energy estimator for both the ON-source and 
background regions, the background histogram was subtracted from 
the ON-source histogram. Subsequently, an energy spectrum was fitted to 
this ``excess histogram'' using the forward folding approach (see e.g.\ 
\citet{fenimore:82}), making use of a simulated set of \gray showers.
The simulated set of showers consisted of 67,500 showers simulated over an area 
$A_{\rm MC}\,=$ $\pi (400\,\rm m)^2$ in the energy range from
50 GeV to 25.6 TeV over nine energy intervals.  The first energy
interval went from 50 to 100 GeV, the second from 100 to 200 GeV and so forth.
In the  $i^{\rm th}$ energy interval, showers were 
simulated according to a power law distribution:
\begin{equation}
\frac{dN^{(i)}_{\rm MC}}{dE}\,=N_{i}\,\times\,(E/\mathrm{1 TeV})^{-\Gamma_{\rm MC}} 
\label{pl}
\end{equation}
with $\Gamma_{\rm MC}\,=$ 2.5. 
Simple integration of Eq.\ \ref{pl} gives the normalization constant 
$N_i$ as function of the lower bound 
$E_{\rm min,i}$ and upper bound $E_{\rm max,i}$ of the $i^{\rm th}$ energy 
interval and the number $n_i$ of showers simulated in that energy
interval:
\begin{equation}
N_{i}\,=\,\frac{-n_i \, \times \, (-\Gamma_{\rm MC}+1)}
{E_{\rm min,i}^{\, \, \, -\Gamma_{\rm MC}+1}-E_{\rm max,i}^{\, \,
    \, -\Gamma_{\rm MC}+1}}
\label{e2}
\end{equation}

We fit the data with two models, a power law model and a power law model 
with an exponential high-energy cutoff:
\begin{equation}
\frac{dN_{\gamma}}{dE}=N_0\times (E/{\rm 1\,TeV})^{-\Gamma}
\end{equation}
and 
\begin{equation}
\frac{dN_{\gamma}}{dE}=N_0\times (E/{\rm
  1\,TeV})^{-\Gamma}\times\exp{(-E/E_0)} \,\,\, ,
\end{equation}
where $N_0$ is the flux normalization at 1~TeV, $\Gamma$ is the photon index,
and $E_0$ is the high-energy cutoff.
For each trial parameter set ${\cal P}$ 
(with ${\cal P}\,=\,\left\{N_{0},\Gamma\right\}$ or
${\cal P}\,=\,\left\{N_{0},\Gamma,E_{0}\right\}$),
another histogram is filled with weighted \gray showers 
that pass the \gray selection cuts.
The weights are computed according to:
\begin{equation}
W_{i}(E;{\cal P})=\frac{\frac{dN_{\gamma}}{dE}(E;{\cal P})}
{\frac{dN^{(i)}_{\rm MC}}{dE}\times\left(A_{MC}\times \Delta t \right)^{-1}} \,\,\, ,
\label{e1}
\end{equation}
where $\Delta$t is the observation time, and $A_{MC}$ is the
area over which showers were simulated.
The numerator in Eq. \ref{e1} gives the model flux for the 
parameter combination ${\cal P}$ at energy $E$. The denominator gives the 
simulated flux. While the weights depend on the true energy of the 
simulated \grays, the showers are filled into the histogram 
according to their reconstructed energy. 
The weighting saves computational time in the fitting procedure, 
as only the weights have to be re-computed for each set of model parameters.

\paragraph{}
We performed a search in parameter space until the parameter combination
$\cal{P}_{\rm min}$ is found that minimizes the $\chi^2$-difference 
between the observed and simulated histograms.
We determined the 1 $\sigma$ error region from the condition \citep{NumRec}
\begin{equation}
\chi^{2}({\cal P}) \leq \chi^{2}({\cal P}_{\rm min})+1  \,\,\, .
\end{equation}

The best fit model parameters and the associated errors are the main
results of a spectral analysis. Plotting individual data points in 
an energy spectrum is well known to be an ill-defined problem. 
A very good discussion can be found in \citet{loredo:89}.
Owing to the finite energy resolution of the telescopes, some information 
about the true energy spectrum is irrevocably lost.
We have experimented with ``deconvolution methods'', as for 
example, the Backus-Gilbert method
\citep{backus:70,loredo:89}. Owing to a combination of almost Gaussian 
distributions of $\ln{(E_{\rm true}-E_{\rm rec})}$ and  
$E_{\rm true}-E_{\rm rec}$ and the modest signal-to-noise ratios
of the TeV \gray energy spectra, we find that deconvolution methods 
improve only very little the effective energy resolution.

We thus use the simplest method to plot flux estimates, based on the
counts in the excess histogram (see e.g. \citet{fenimore:82}). 
For an energy bin stretching from 
$E_1$ to $E_2$, we plot the flux value at the energy 
$E =$ 10$^{(\log{(E_1)}+\log{(E_2)})/2}$.
The flux value is given by scaling the best-fit model according to the
observed number of excess counts:
\begin{equation}
f = \frac{dN}{dE}\left({\cal P}_{\rm min}\right) \times
\frac{k_i}{<k_i>} \,\,\, .
\end{equation}
Here, $k_i$ is the number of excess counts in the i$^{\rm th}$ bin
of the signal histogram, and $<k_i>$ is the sum of weights of the simulated 
events in the i$^{\rm th}$ bin.

\begin{figure}
  \epsscale{1.0}
  \plotone{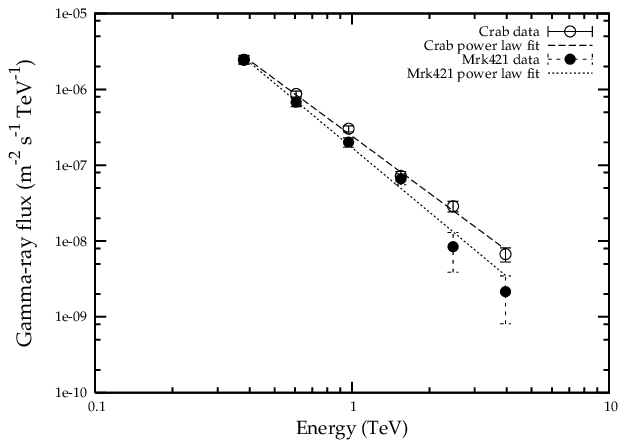}
  \caption{\label{spectra}Whipple TeV spectrum of the Crab Nebula and Mrk 421.
    The dashed and dotted lines give the results of power law fits for
    the Crab and Mrk~421, respectively.}
\end{figure}

\begin{figure}
  \epsscale{0.80}
  \plotone{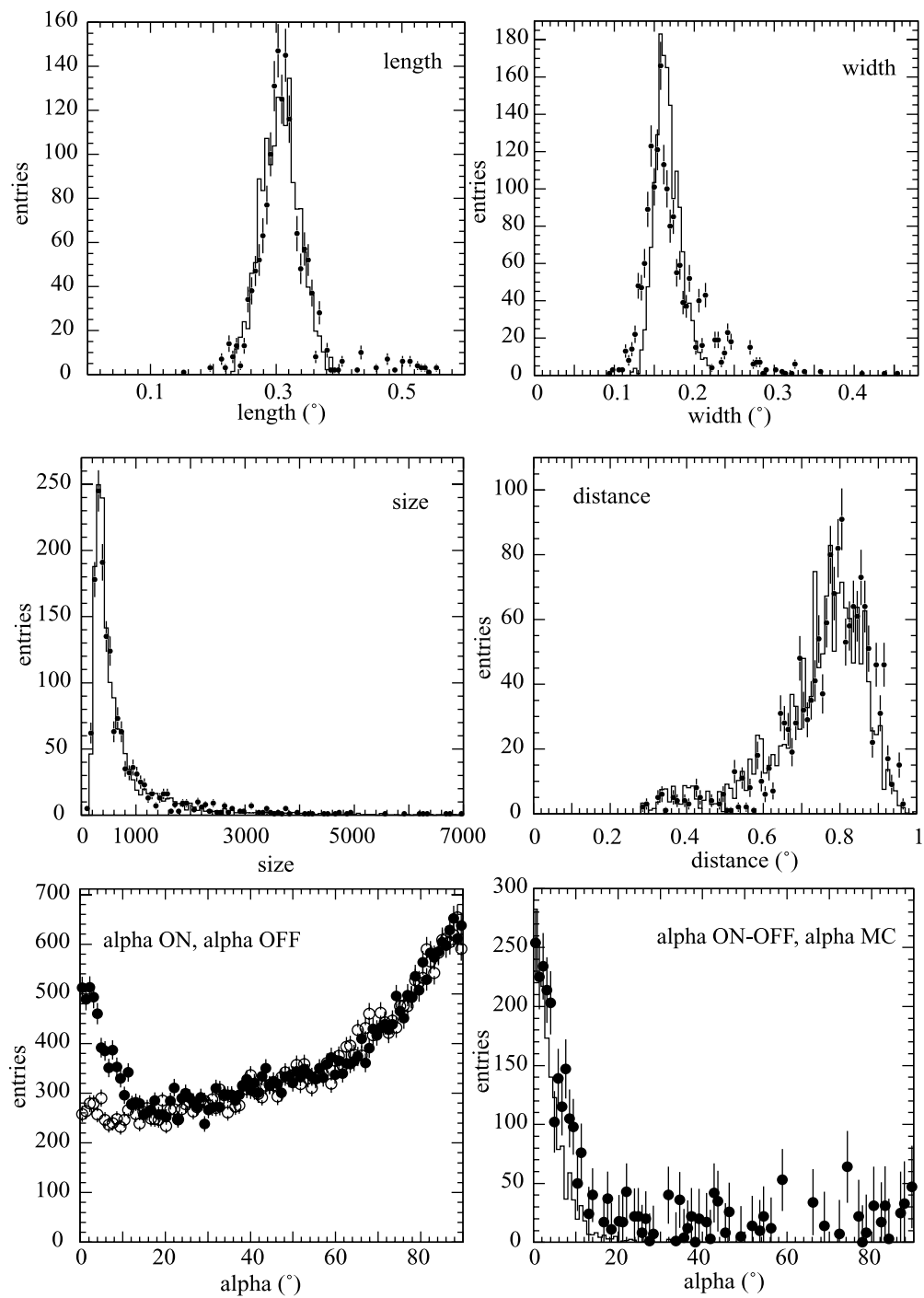}
  \caption{\label{comparison} Hillas parameter distributions for
    Whipple 2002-2003 Crab ON/OFF data and simulated data. Histograms show simulated data,
    while data points show Crab data.}
\end{figure}

Figure \ref{spectra} shows the Crab spectrum from small zenith angle data
($<$ 30$^\circ$) acquired between 14 September 2002 (UT) and 24
March 2003 (UT), and the time averaged TeV 
\gray spectrum of Mrk 421 for the data set of the multiwavelength
campaign.
Both data sets are from the Whipple 10 m telescope.
The power law fit to the Crab data gives a flux normalization
of $2.42 \pm 0.11$ $\times$ 10$^{-11}$ photons cm$^{-2}$ s$^{-1}$ TeV$^{-1}$,
and a photon index of $\Gamma\,=$ 2.5 $\pm$ 0.1. 
The $\chi^2$ is 5.27 for four degrees of freedom.
These results should be compared to previous results.
The Whipple collaboration obtained a flux normalization
$N_0\,$= (3.25$\pm$0.14$\pm$0.60)
$\times$10$^{-11}$ photons cm$^{-2}$ s$^{-1}$, and a spectral index
$\Gamma\,=$ 2.49$\pm$0.06$\pm$0.04 \citep{Hill:98}.
The HEGRA collaboration published
$N_0\,$= (2.83$\pm$0.04$\pm$0.60)
$\times$10$^{-11}$ photons cm$^{-2}$ s$^{-1}$, and 
$\Gamma\,=$ 2.62$\pm$0.02$\pm$0.05 \citep{Ahar:04}.
Finally, the CAT collaboration obtained
$N_0\,=$ (2.21 $\pm$ 0.05 $\pm$ 0.60)
$\times$10$^{-11}$ photons cm$^{-2}$ s$^{-1}$, and 
$\Gamma\,=$ 2.80$\pm$0.03 $\pm$ 0.06 \citep{Piro:01}.
The first error values are the statistical error values, and the second
are the systematic error values (both errors on $1\sigma$ confidence level).
The systematic errors on the absolute flux are about 25\% and derive
from the uncertainty of the energy threshold of the instruments combined
with the steep spectrum of the Crab Nebula.
Our result derived here lies below the previously published values 
of Whipple and HEGRA and above the value published by CAT.
The discrepancies are comparable to the one-sigma confidence levels.
It should be noted that the Whipple results published in 1998 were taken
with a significantly different hardware configuration. Furthermore,
we relied here on a energy threshold calibration with muons, while
\citet{Hill:98} used the comparison between the detection
rate of simulated and observed Cosmic Rays to calibrate the
energy threshold of the telescope. 
Each calibration method has its own systematic uncertainty, and it is
difficult to decide which one is more reliable.
The discussion shows that one should consider the full systematic error
when comparing the two Whipple results with each other. 
The three experiments quote a systematic error of $\simeq$0.05 on the
photon index. Here, our result agrees well with the previous Whipple 
measurements, the HEGRA spectrum is somewhat steeper and the CAT 
spectrum is significantly steeper. The comparison of all the four results
shows that systematic errors are larger than estimated.
In the case of the blazar observations discussed below, 
the uncertainty of the absolute energy threshold are not that important,
as we are mostly interested in relative flux variations.
We correct our fluxes with a ``throughput factor'' derived
from the Cosmic Ray detection rate measured during each data run
to correct for variations in the atmospheric conditions.
As Cosmic Ray showers are not identical to air showers, 
we estimate that the systematic uncertainty on diurnal 
fluxes is 10\%, and the systematic error on diurnal 
photon indices is 0.1.
\paragraph{}
The power law fit to the Mrk 421 data gives a flux normalization
of $1.7 \pm 0.1$ $\times$ 10$^{-11}$ photons cm$^{-2}$ s$^{-1}$ TeV$^{-1}$,
and a photon index of $\Gamma\,=$ 2.8$\pm$0.1. 
The $\chi^2$ is 5.06 for 4 degrees of freedom.
The flux and photon index lie in the range of previous observations 
\citep{Zwee:97,Ahar:99,Kren:99,Kren:03,KrenDwek:03}.  
We find that the photon statistics
do not allow us to derive meaningful constraints on the high-energy
cutoff $\mathrm{E_{0}}$.
\paragraph{}
Using the best-fit parameters from the Crab spectral analysis, we can weight the 
Monte Carlo events by the determined spectrum and compare several simulated 
distributions of the image parameters with the background subtracted distributions 
for the Crab data (Fig.\ \ref{comparison}). The good agreement between 
simulated and experimental data verifies that the simulations describe 
the air showers and the detector response adequately.

\section{Results from the Multiwavelength Campaign}
\subsection*{Overview}
\begin{figure}
  \epsscale{0.50}
  \plotone{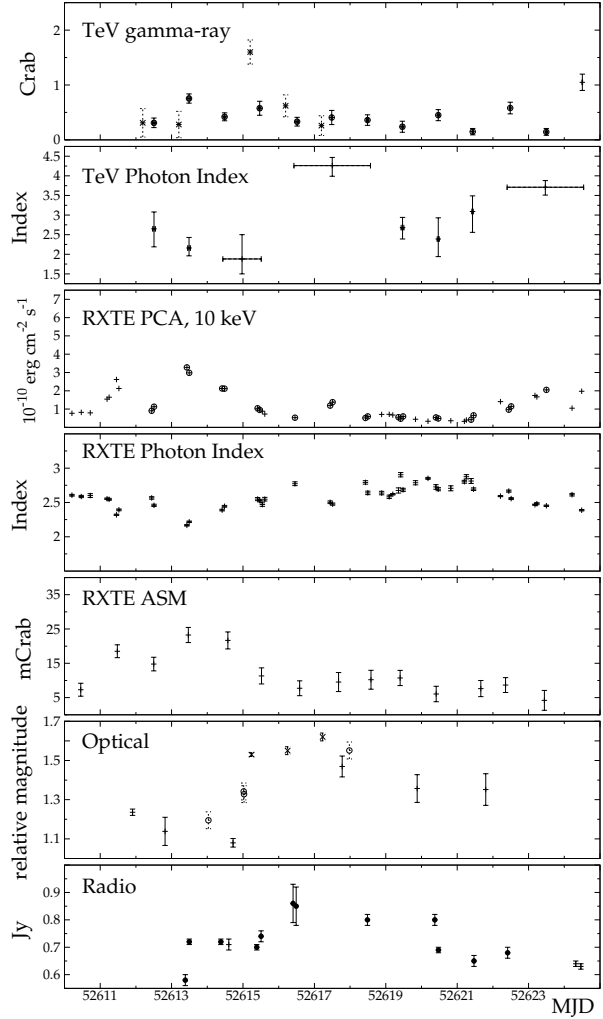}
  \caption{\label{lc1}Multi-wavelength data from December, 2002. 
    The \gray data points show per-night averages, in Crab units. Starred data points
    signify data taken from the HEGRA CT1 telescope, while crosses denote Whipple 10m data.  
    The error bars on the RXTE PCA data are not shown as they are
    smaller than the symbol size, and have units of $10^{-10}$ erg cm$^{-2}$ s$^{-1}$ at 10 keV.
    The circled X-ray and \gray data points overlapped or were taken less than 5 min apart.
    The TeV \gray and RXTE photon indices show $\Gamma$, where $dN/dE \propto E^{-\Gamma}$. 
    the ASM data are given in mCrab.
    In the optical band, open circles show the WIYN V band data,
    crosses show the Boltwood R band data, and 'x' denotes La Palma R
    band data.  All of the optical data are in relative magnitude units.
    The open (filled) circles in the radio band show measurements that overlapped or were
    taken within 5 min of a TeV \gray observation (X-ray and TeV \gray observation).
    The radio data are given in Janskys.}
\end{figure}

\begin{figure}
  \epsscale{0.50}
  \plotone{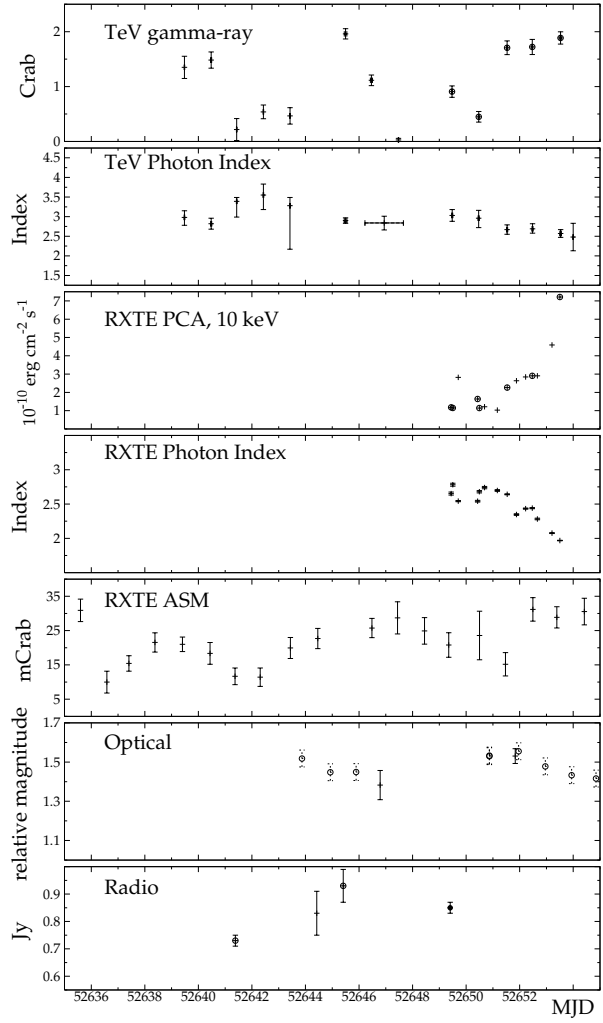}
  \caption{\label{lc2}Same as Fig.\ \ref{lc1} for the data from January 2003.}
\end{figure}

Figures \ref{lc1} and \ref{lc2} combine all the light-curves measured
in December 2002, and January 2003, respectively. From top to bottom, the figures 
show the TeV \gray data, TeV photon indices $\Gamma$, (where $dN/dE \propto E^{-\Gamma}$), 
{\it RXTE} PCU 10 keV flux amplitudes (from the 4-15 keV spectral fits), the 
4-15 keV photon indices, the {\it RXTE} ASM 2-12 keV fluxes, the optical data, 
and the radio data.

The TeV \gray fluxes varied between 0 to $\sim 2$ times that of the
Crab Nebula, 
with slightly higher fluxes observed during the second half of the campaign.
We determined TeV photon indices on a night to night basis whenever the flux was 
sufficiently high to warrant a spectral analysis. For epochs of low fluxes
(December 6 and 7, 2002 (UT) (MJD 52614-52615), 
December 8, 9 and 10, 2002 (UT) (MJD 52616-52618), 
December 14, 15, and 16, 2002 (UT) (MJD 52622-52624), and January
7 and 8, 2003 (UT) (MJD 52646-52647)) we combined the data of several 
nights to determine an energy spectrum. A $\chi^{2}$ test of statistical
variability was performed by fitting the entire TeV photon index
dataset to a constant function.  The best fit to the data, with
$\mathrm{2} \sigma$ errors, is $\Gamma = -2.864 \pm 0.097$, with a
$\chi^{2}$ value of 46.5 for 20 degrees of freedom. The probability of
obtaining this value by chance is $2.2\times 10^{-4}$.  Seven data
points lie outside the $\mathrm{2} \sigma$ confidence region.

The 4-15 keV {\it RXTE} data is shown in Figs.\ \ref{lc1} and \ref{lc2}. 
The X-ray fluxes varied between $0.2\times 10^{-3}$ cts cm$^{-2}$
s$^{-1}$ keV$^{-1}$ and $4.5\times 10^{-3}$ cts cm$^{-2}$ s$^{-1}$ keV$^{-1}$.
Strong flares were observed on December 3, 2002 (MJD 52611), 
December 5, 2002 (MJD 52613), January 10, 2003 (MJD 52649), 
and on January 14, 2003 (MJD 52653).  The 4-15 keV X-ray photon
indices show a large range of values from 
$\Gamma= 1.97$ to $\Gamma= 2.90$. 

The values of $\Gamma \ge \sim  2$ indicate that the presumed synchrotron
SED peaked at and below the 4-15 keV energy range covered by observations.
In two cases, the X-ray index varies very rapidly:
on MJD 52619 it changes by $\Delta\Gamma\,= 0.22$  within 1.4 hrs, and on 
MJD 52650 it changes by $\Delta\Gamma\,= 0.14$ within 1.6 hrs.


Over the duration of the campaign, the 4-15 keV photon index changed by 
$\Delta\Gamma\,\approx$ 1. Synchrotron cooling of a power law distribution 
of electrons produces a spectral break equal to, or smaller than 
$\Delta\Gamma\,=$ 0.5 if they high-energy cutoff of the electron spectrum 
is outside the range sampled by the observations. If the power-law index 
of the accelerated particles does not change with time, the detection of spectral 
variations with $\Delta\Gamma>$ 0.5 thus implies that the 4-15 keV fluxes are 
influenced by the high-energy cutoff of the relativistic electron population. 
Alternatively, flares might be associated with changes of several
parameters constraining the emitting plasma, as for example the jet 
magnetic field or the jet beaming angle.

\paragraph{}
The {\it RXTE} PCU and ASM fluxes seem to trace each other, although the sparse
sampling of the PCU data and the large statistical error bars of the ASM data
do not allow us to draw definitive conclusions.
The optical and radio data show substantial variability by 
about $\pm$50\% of the mean flux, but no detailed correlation is
present.  This is not a suprise since for the optical data one expects
sizeable contributions from the galaxy light and perhaps thermal
emission from the accretion disk in addition to optical
synchrotron. Furthermore, one expects longer variability timescales
for the lower energy synchrotron radiation, since the lower energy
electrons which produce this radiation cool more slowly. 
\subsection*{The X-Ray/TeV \gray Flux Correlation}
We studied the X-ray and TeV \gray flux correlation with the help of the
discrete correlation function (DCF) of \citet{edelson}.
The DCF gives the linear correlation coefficient for the two light curves 
as a function of a time lag between them. The DCF is the standard tool used in 
the case of sparsely sampled data and gives fewer spurious results than 
a traditional correlation function analysis that interpolates between 
light-curve data points. We determined the statistical significance of the 
correlation coefficients with the help of a set of simulated \gray light 
curves, computing for each simulated light curve the DCF with the 
observed X-ray data \citep{buckley04}. These light-curves are generated 
by a superposition of triangular shots, with all shots having the same amplitude 
and the same rise and fall time. 

\begin{figure}
  \epsscale{1.0}
  \plotone{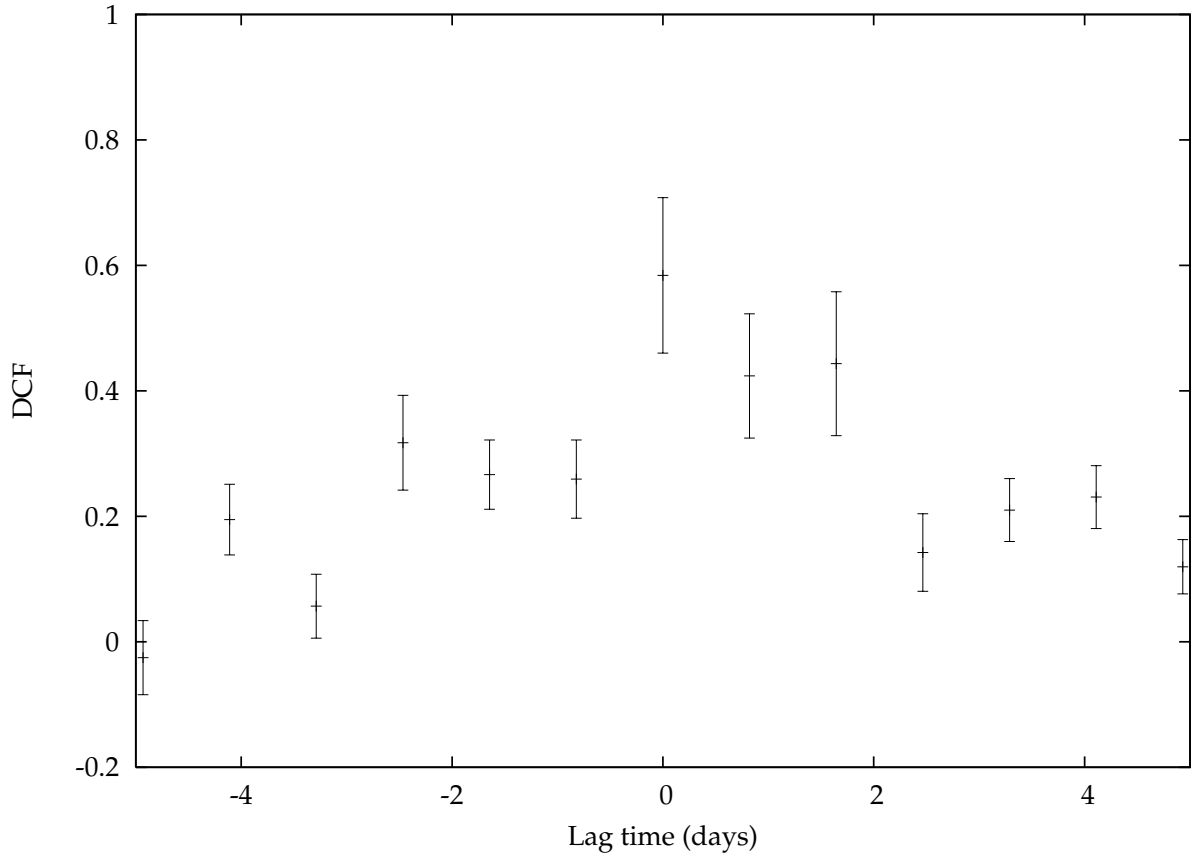}
  \caption{\label{dcf}Discrete correlation function of the complete
    X-ray and \gray data set. A positive time lag means the \gray flux 
    precedes the X-ray flux.}
\end{figure}

\begin{figure}
  \epsscale{1.0}
  \plotone{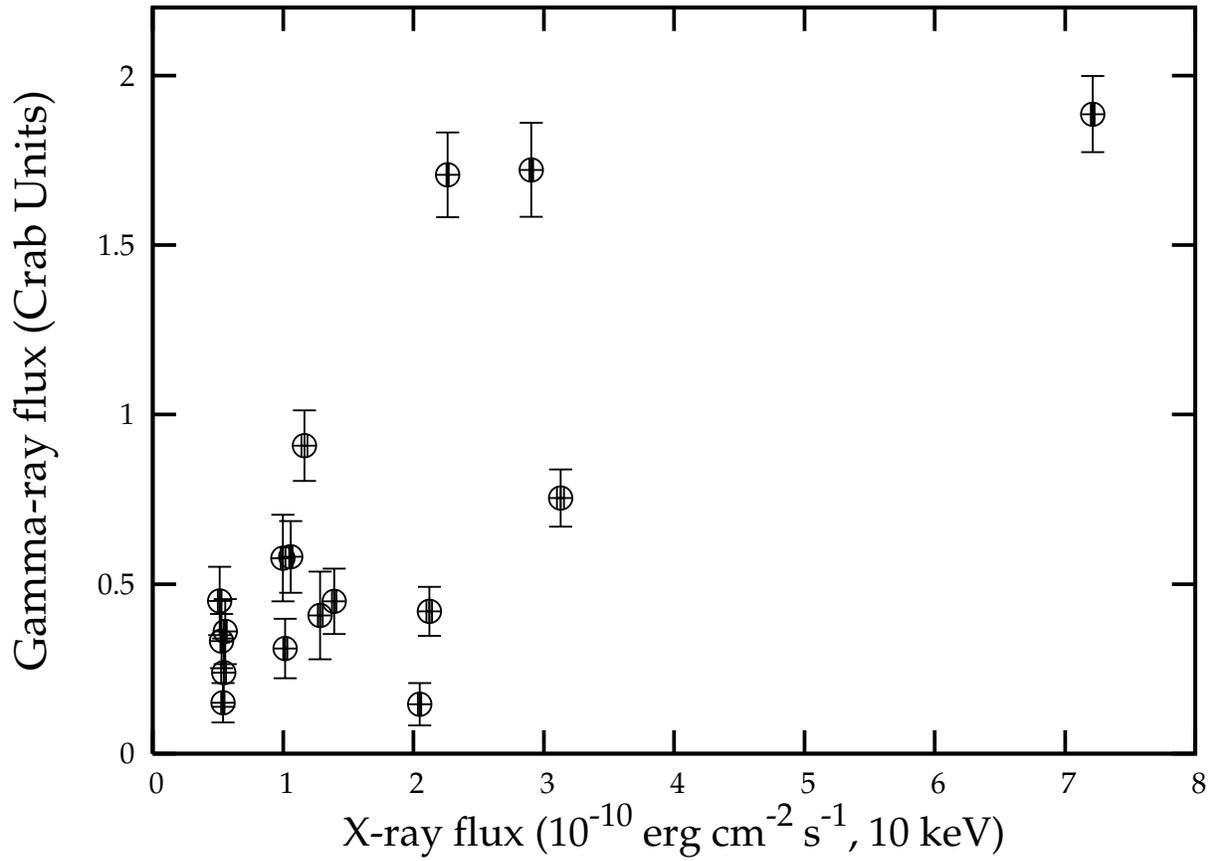}
  \caption{\label{xgamma}Plot of the TeV \gray versus X-ray flux
    correlation for measurements for all overlapping data sets.} 
\end{figure}

Figure \ref{dcf} gives the DCF for the X-ray and TeV \gray data sets.  
For a time lag of zero days, we find a DCF value of $0.58 \pm 0.12$. 
The simulated data sets show that the correlation is significant. For
uncorrelated lightcurves consisting of triangular shots with the same
structure function, we calculate that
the chance probability to get a larger DCF value at a time lag of zero
days is 3.12\%. Figure \ref{xgamma} shows the X-ray/TeV \gray 
flux correlation for all overlapping observations. The measurements entering 
this figure are shown by the circled data points in Figs.\ \ref{lc1}
and \ref{lc2}. 
Each of these overlapping observations are simultaneous to within 5 minutes. 
The correlation shows substantial scatter, with similar X-ray (TeV) 
fluxes sometimes corresponding to TeV (X-ray) fluxes that differ 
by more than a factor of 2 from each other. The scatter may be an 
inherent property of the emission mechanism. Alternatively,
it may be caused by a short time lag between the flares in the two
bands, not resolved by the sparse sampling during the campaign.

\subsection*{Evolution of the X-ray Photon Indices}
\begin{figure}
  \epsscale{1.0}
  \plotone{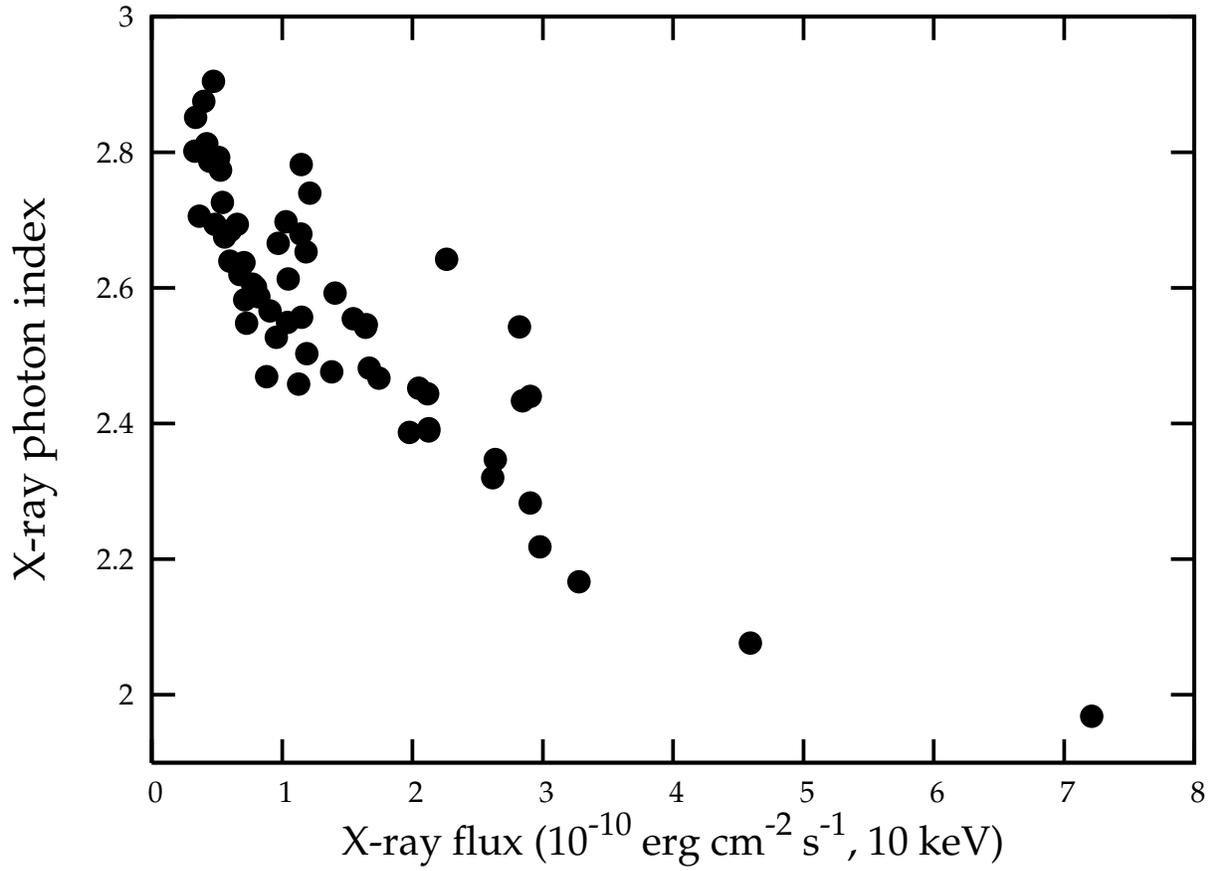}
  \caption{\label{xxcorr}Correlation of the 10 keV X-ray flux and the3-14 keV
    photon index (both: RXTE PCA data).}
\end{figure}

Figure \ref{xxcorr} shows the correlation between the 10 keV fluxes and 
the 4-15 keV photon indices. 
A correlation can clearly be recognized in the sense that higher 
fluxes are accompanied by harder energy spectra. A similar correlation 
(harder energy spectra for higher fluxes) have been reported for 
other BL Lac-type objects, e.g. Mrk 501 \citep{Kata:99,Kraw:00}, 
1ES 1959+650 \citep{Kraw:02}, H~1426+428 \citep{Falc:04}, 
and PKS 2155-304 \citep{kataoka00}.

\begin{figure}
  \epsscale{1.0}
  \plotone{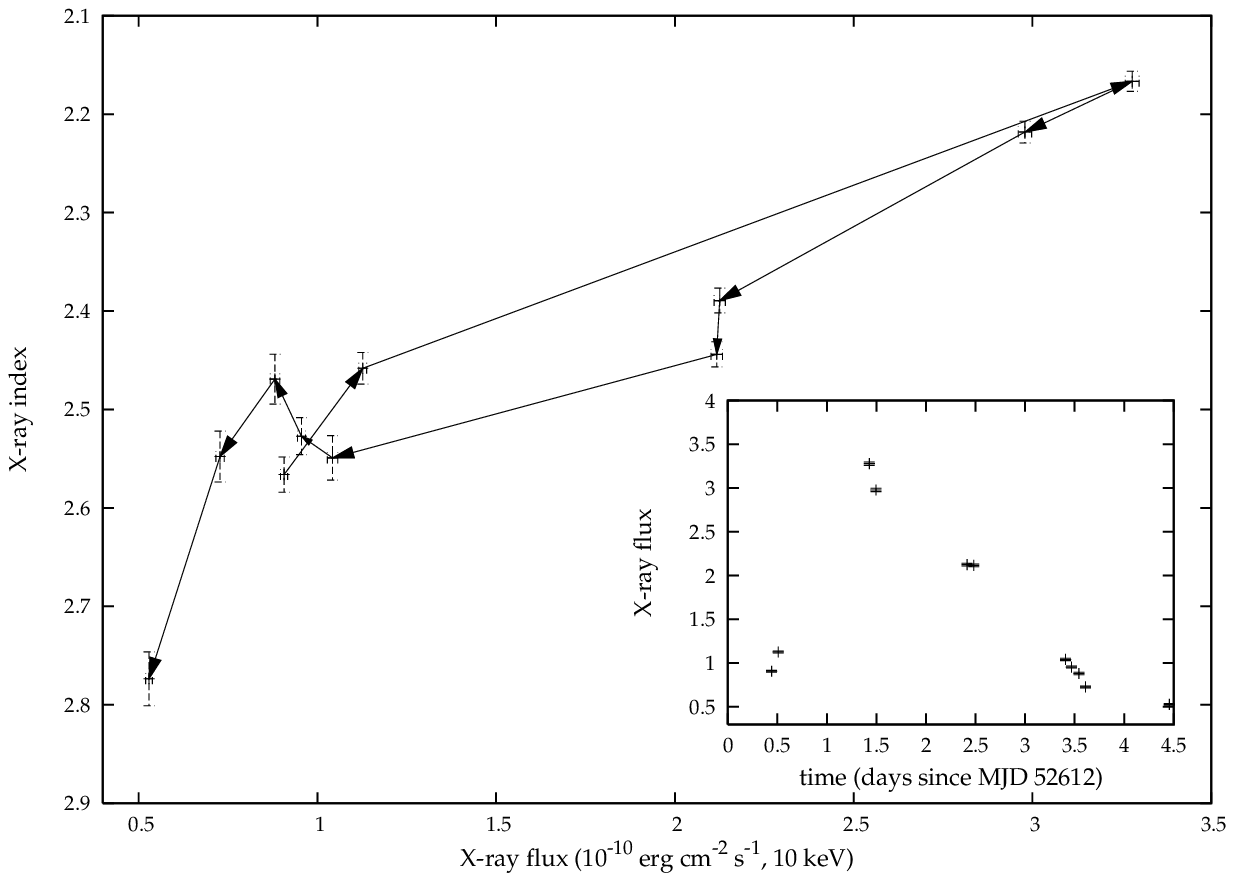}
  \caption{\label{hyst}X-ray power law photon index versus X-ray flux for the MJD
    52612-52615 X-ray flare. }
\end{figure}

\begin{figure}
  \epsscale{1.0}
  \plotone{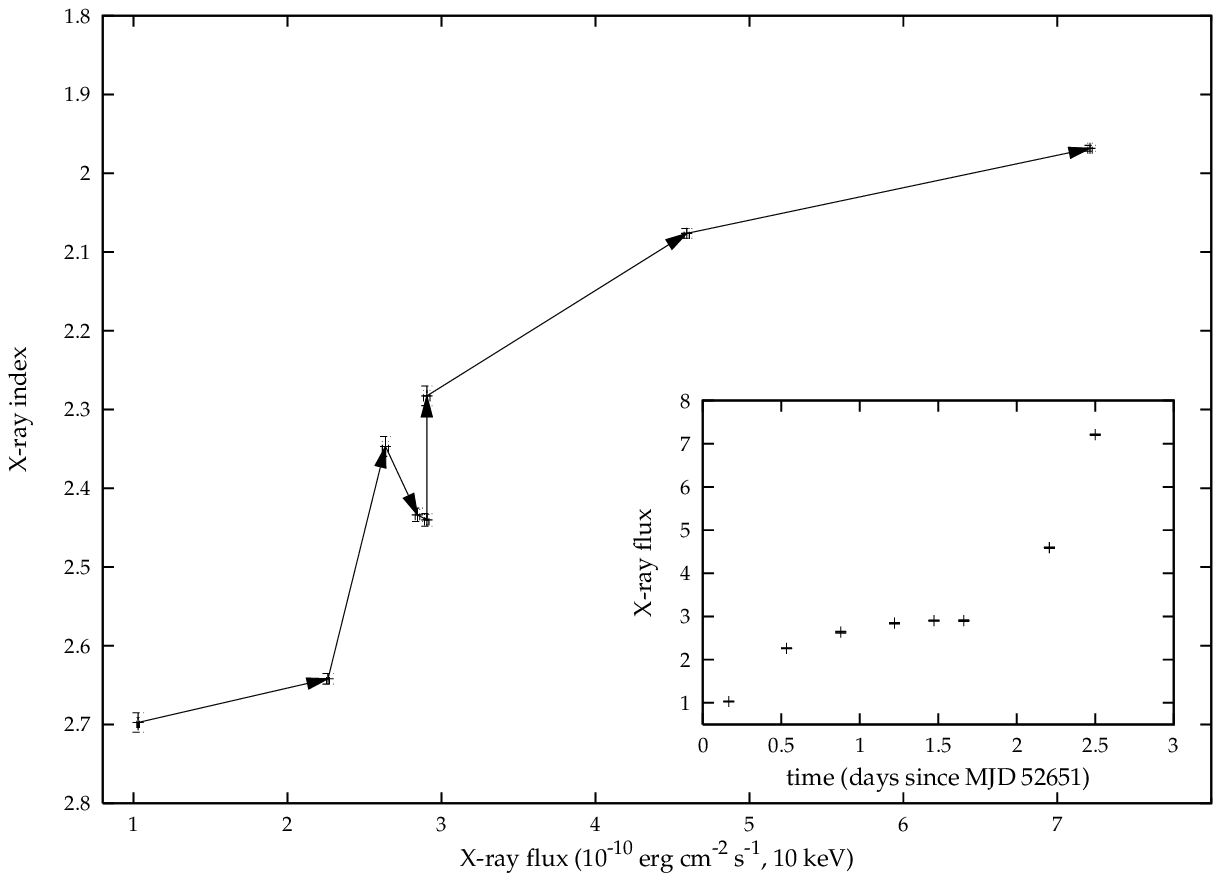}
  \caption{\label{hyst2}X-ray power law photon index versus X-ray flux for the MJD
    52651-52653 X-ray flare. }
\end{figure}

We further scrutinized the temporal evolution of the photon indices with
so called X-ray ``hysteresis'' curves \citep{Taka:96, Kirk:99}, 
plotting the X-ray photon index as a function of X-ray flux, and
indicating the temporal sequence of the measurement throughout the
evolution of individual flares. In the simplest model whereby flares
are formed by short lived shocks (e.g. internal shocks from colliding
blobs), one expects the temporal evolution to be dictated by the
interplay of the acceleration, cooling, and confinement times.
For the Fermi mechanism, the particle energies reached depend on the
allowed acceleration time. 
Thus, during the beginning of flares, the X-ray and \gray 
energy spectra are expected to harden. Once synchrotron cooling starts 
to dominate the particle energy spectra, the emitted photon energy 
spectra are expected to soften.
We studied two X-ray flares, one occurring at the beginning of 
the campaign (MJD 52612 to MJD 52615), and the other at the 
end (MJD 52651 to MJD 52653).  
The first flare is shown in Fig.\ \ref{hyst}.  This flare 
coincided with an increase of the TeV \gray flux by a factor of 
2.4 from MJD 52612 to MJD 52613. The X-ray spectrum hardens during 
the rising phase of the flare and softens during the
decaying phase. Furthermore, the X-ray spectrum is softer during the falling 
phase than during the rising phase. 
The ``clockwise'' evolution in the $\Gamma_{x} - F_{x}$ plane is consistent 
with the expectations from stochastic Fermi acceleration and 
synchrotron cooling as described above.
The data of the second flare are shown in Fig.\ \ref{hyst2}.
The TeV \gray flux increased from MJD 52650 to MJD 52651 by a factor of 3.8 and remained
roughly constant during the following two nights of observations.
While the X-ray flux increased from MJD 52652 to 52653 by a factor of $\sim 2.5$,
the TeV flux measured at the same time as the X-ray fluxes did not 
increase substantially.
The general trend is that the spectrum hardens as the flux increases, although
at MJD 52652.25, the spectrum softens temporarily while the flux is still increasing.
Unfortunately, our observations did not cover the decaying phase of the flare.
\section{Discussion}\label{discussion}
\paragraph{}
The multiwavelength campaign showed Mrk 421 in a level of intermediate activity.
During the observational campaign, Mrk 421 showed significant flux
variability in the radio, optical, X-ray and \gray bands and
significant spectral variability at X-rays and TeV \grays. While
we measured an average TeV \gray photon index of $\Gamma\,=$ 2.8, the observations 
revealed evidence for spectral variability on a time scale of days. In
particular, the data suggest very soft energy spectra with
$\Gamma \approx 4$ during the first half of the observation
campaign. One of the most interesting results from this campaign is that 
the X-ray and TeV \gray fluxes are correlated on the $\sim $97\% confidence level, 
but that we find widely different TeV \gray fluxes for a single
X-ray flux and vice versa. 
The most extreme example is the ``orphan X-ray flare'', seen on 
January 13, 2003 (MJD 52653). The loose X-ray/TeV
correlation may suggest that the model parameters (e.g. the volume of
the emission zone) change with time, or that the commonly made
assumption of a single synchrotron self-Compton (SSC) emission 
zone is an over-simplification.
Our previous observations of Mrk 421 had already shown a rather loose 
X-Ray/TeV \gray correlation \citep{Blaz:05} and the same applied for
1ES 1959+650 \citep{Kraw:04}. In the case of Mrk 501, a rather tight 
correlation has been reported \citep{Kraw:00,Kraw:02}.
\paragraph{}
The analysis of the correlation between the X-ray flux and photon index
during a flare indicated a ``clockwise'' hysteresis. For Mrk 421, 
\citet{Taka:96} also reported clockwise evolution.
However, \citet{Taka:00} reported evidence for both, 
clockwise evolution during some flares and anti-clockwise evolution 
during other flares. If the SSC model indeed applies, these results may 
imply that the characteristic times scales of the most important processes 
(acceleration time, radiative cooling time, escape time) change from flare 
to flare and thus yield the different observed hysteresis behaviours.
Recently, \citet{Soko:04} emphasized that the geometry of the 
emitting region and its orientation relative to the line of sight 
influences the observed flux and spectral evolution and might thus
further complicate the interpretation of the results.

The X-ray and TeV \gray emission from Mrk 421 data have been modeled with
synchrotron Self-Compton codes by many groups, see e.g. 
\citep{Inou:96,Bed:97,Bed:99,Boet:97,Mast:97,Taka:01,Kraw:01,Kono:03,Kino:02,Blaz:05}.
A crucial model parameter is the jet Doppler factor $\delta_{\rm j}$.
The published models with Doppler factors $\delta_{\rm j}$ of 20 or lower generally 
predict TeV energy spectra that are softer than the observed ones, especially 
if a correction for extragalactic absorption would be applied which steepens the
energy spectra considerably. Models with Doppler factors $\delta_{\rm j}$ on the 
order of 50 give satisfactory model fits to both the X-ray (synchrotron) and 
the TeV (Inverse Compton) emission (see the self-consistent modeling of 
\citet{Kraw:01,Kono:03} and the discussions by \citet{Tave:04,Pine:05}).
\citet{Pine:05} observed the Mrk 421 parsec-scale radio jet 
with the Very Large Baseline Array (VLBA). Remarkably, they find apparent pattern 
speeds of only $\sim 0.1 \mathrm{c}$. As discussed by the authors, the highly 
relativistic motion inferred from TeV observations can be reconciled with 
the modestly relativistic flow calculated from VLBA observations by postulating 
that the jet slows down between the sub-parsec (TeV) and parsec (VLBA) regimes.
%
It may be possible to describe the multi-wavelength data with a synchrotron-Compton 
model and lower Doppler factors by invoking additional seed photons. 
Two new model variants that combine ingredients of SSC and external Compton 
models have been proposed by \citet{Geor:03} and by \citet{Ghis:05}. 
While the first authors assume that downstream emission regions provide seed 
photons, the second authors speculate that the jet is
composed of a fast spine with a slow-moving envelope. In this model,
the fast spine emits the X-ray and \gray radiation. 
Modeling of the data with these two inhomogeneous models is outside 
the scope of this paper.

\begin{figure}
  \epsscale{1.0}
  \plotone{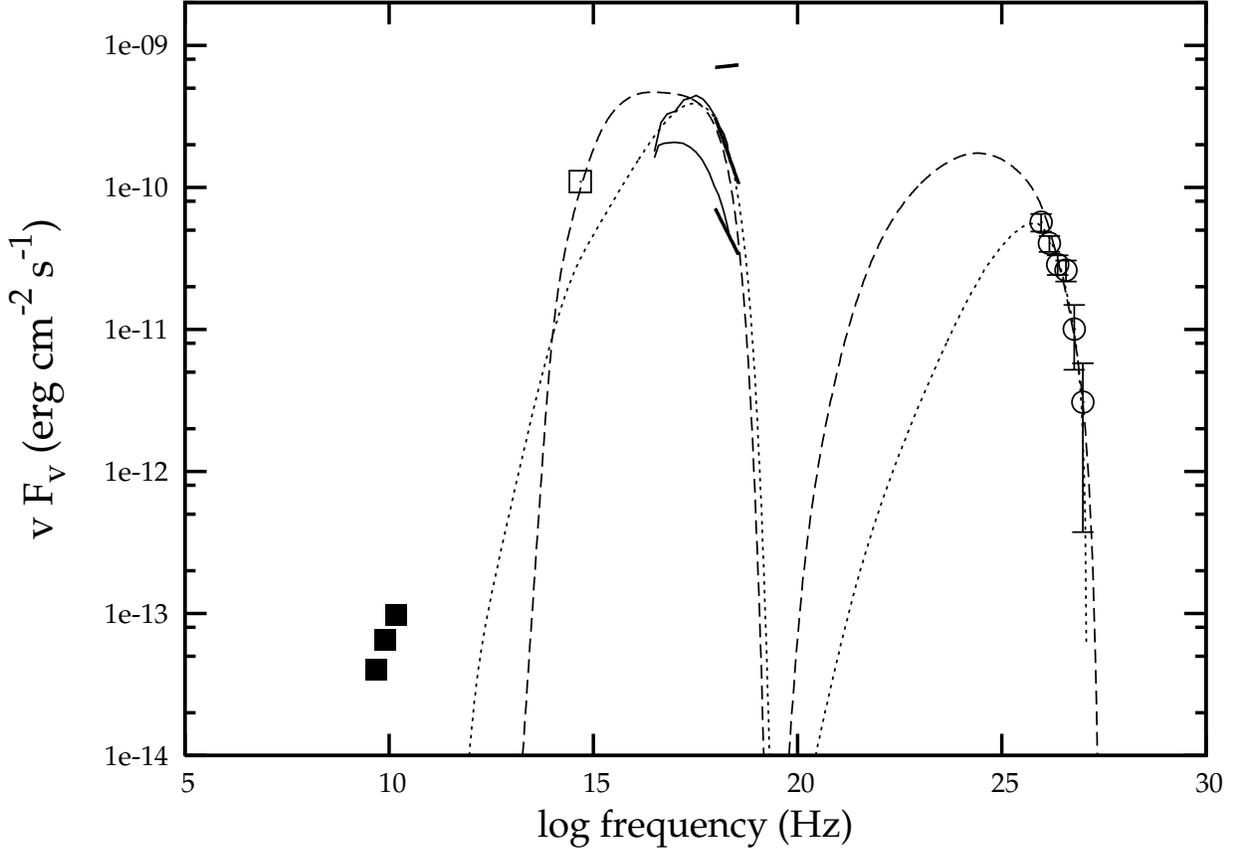}
  \caption{\label{nufnu}Mrk 421 Spectral Energy Distributions measured 
    during the campaign. The data points show the radio to \gray data. 
    The radio and optical spectra show the average fluxes observed during 
    the campaign. At X-rays we show three spectra, one
    from the {\it RXTE} pointing with the highest flux observed during the
    campaign, one from the pointing with the lowest flux, and one 
    spectrum at intermediate flux levels.
    The intermediate X-ray spectrum and the gamma-ray spectrum were determined
    using for both only the data taken during nights with simultaneous X-ray 
    and gamma-ray observations.
    The solid curved lines show, for comparison, a low-flux and a high-flux 
    energy spectrum measured with {\it BeppoSAX} (``Beppo'' Satellite 
    per Astronomia X) during the 1998 flaring period \citep{Foss:00}. 
    the long dashed lines show the results from a simple Synchrotron
    Self-Compton model with $\delta = 1000$, while the short dashed lines show the results 
	with $\delta = 50$. }
\end{figure}

In this discussion we do not want to embark on comprehensive modeling
of the data from the entire campaign. Our main aim is to draw the attention 
of the reader to a single remarkable fact: while it is difficult to model 
the data with Doppler factors on the order of 20 and lower, much higher 
Doppler factors cannot be excluded right away.
Fig.\ \ref{nufnu} shows two synchrotron self-Compton models based on the simple
snapshot code of  \citet{Kraw:04}. The code assumes a single spherical emission 
volume of radius $R$ relativistically approaching the observer with a jet 
Doppler factor $\delta_{\rm j}$. The emission volume is homogeneously filled 
with a tangled magnetic field of strength $B$ and a non-thermal electron population.
The electron energy spectrum follows $dN/d\gamma\propto$ $\gamma^{-p}$ with 
$p\,=$ 2 for electron Lorentz factor $\gamma$ between $\gamma_{\rm min}$ and  
$\gamma_{\rm b}$ and $p\,=3$ for Lorentz factors between $\gamma_{\rm b}$ 
and $\gamma_{\rm max}$.
The code models extragalactic absorption owing to the 
$\gamma_{\rm TeV} + \gamma_{\rm CIB} \rightarrow e^+ e^-$ 
pair-production processes of TeV photons on photons from the 
Cosmic Infrared Background (CIB) using the CIB model of \citet{Knei:02}. 

We discuss two models. We show the first model for illustrative purposes only.
It uses the "conventional" model parameters ($\delta_{\rm j}\,= 50$) derived from 
the time-dependent self-consistent analysis of a different but similar data set \citep{Kraw:01}.
The second model uses a very high Doppler factor ($\delta_{\rm j}\,= 1000$).
All the model parameters are given in Table \ref{fits}. 
For both models, we assured that the model parameters were 
chosen self-consistently. Causality arguments require that the 
radius $R$ satisfies $R<\delta_{\rm j}c\Delta T_{\rm obs}\,=\,2.7\times 10^{15}$~cm
for $\delta_{\rm j}\,=50$ and $R<5.4\times10^{16}$~cm for $\delta_{\rm j}\,=1000$
for a flux variability time scale of $\Delta T_{\rm obs}\,=$ 30~min.
Note that the flux variability time scale sets a lower limit
on $R$ but no upper limit, if the flux variability time scale is
not dominated by light travel time effects but by other effects
(e.g.\ by the stability of a strong shock front).
We checked that the SED (i) fits the X-ray and TeV
\gray data, and (ii) is consistent with the expected
spectral shape owing to radiative cooling.
In the first model, the latter self-consistency is assured by
our previous self-consistent time-dependent modeling.
In the second model, we construct an electron spectrum based on the
general results for electrons suffering synchrotron and
Inverse Compton losses \citep{Syro:59,Kard:62,Ginz:64,Pach:70,Inou:96}).
We assume that the electron energy spectrum breaks at 
$\gamma_{\rm b}\,=$ $1.8\times 10^3$.
The laboratory-frame synchrotron cooling time for electrons at the break
is $t_{\rm s}\,=$ $[\frac{4}{3}$ $\sigma_{\mbox{\small T}}\,$ $c\,$ $\delta_{\rm j}\,$
$\frac{B^2}{8\pi\,m_{\rm e}\,c^{2}}\,$ $\gamma_{\rm b}$ $]^{-1}~$
$\approx\, 28min $ ($\sigma_{\mbox{\tiny T}}$ is the Thomson cross section and
$m_{\rm e}$ is the electron mass).
An electron spectrum as the one used here could result from the
radiative cooling of a $p\,=2$ electron energy spectrum that extends from
$\gamma_{\rm min}$ to $\gamma_{\rm max}$, resulting in a
spectrum with $p\,=2$ and $p\,=3$  below and above $\gamma_{\rm b}$,
respectively. 

As shown in Fig.\ \ref{nufnu}, both models fail to predict the observed radio fluxes as
a consequence of synchrotron self-absorption. We would like to emphasize that we do not 
regard the discrepancy as a shortcoming of the model.
Electrons producing the radio emission cool on much longer time scales, and the 
radio emission will be dominated by an accumulation of downstream plasma which 
stopped contributing to the X-ray and TeV emission long time ago.
We could model the radio emission with another emission component (see e.g.\ 
\citet{Blaz:05}). However, doing so is arbitrary: for a small number
of data points we would add an additional model component with many 
free model parameters. 

In Table \ref{fits} we list for both models the magnetic 
field energy density $u_{\rm B}$, the energy density in electrons
$u_{\rm e}$, the ratio $r = u_{\rm e}/u_{\rm B}$, and the kinetic 
luminosity $L_{\rm k} = \pi R^{2} c \Gamma^{2} (u_{\rm e} + u_{\rm B})$ for
$\Gamma\,=$ $\delta_{\rm j}$ \citep{Bege:94}. The model with a low Doppler 
factor is closer to equipartition between magnetic field and particles.
The kinetic luminosity is similar for the two models, with a high 
radiative efficiency of the low-$\delta_{\rm j}$ model making up for the 
stronger boosting of the high-$\delta_{\rm j}$ model. 

If taken seriously, the model with $\delta_{\rm j}\,=$ 1000 would imply that the 
X-ray and TeV \gray emission is produced by an ultra-relativistic 
particle dominated wind, very close to the supermassive black hole. 
The fact that seven blazars have been detected at TeV energies seems to 
argue against extremely relativistic outflows with bulk Lorentz 
factors $\Gamma$ on the order of 1000, as isotropic emission would 
be beamed into an opening angle of $\Gamma^{-1}$ and would make the 
observation of the emission unlikely. However, the argument only 
applies if the jet opening angle is equal or smaller than $\Gamma^{-1}$.
Having a larger jet opening angle would require a higher total 
jet-luminosity as some jet segments would not contribute to the 
observed emission. However, the jet-luminosities listed in the table 
are several orders of magnitude below the Eddington luminosity
of a $\sim 10^{8.4}$ solar mass black hole that is suspected to be at the core
of Mrk 421 \citep{Bart:03,Falo:02}.

\acknowledgements
We acknowledge an anonymous referee for helpful comments.
We acknowledge the technical assistance of E. Roache and J. Melnick.  
This research is supported by grants from the U.S. Department of
Energy, the National Science Foundation, the Smithsonian
Institution, by NSERC in Canada, by Science Foundation Ireland, 
and by PPARC in the UK. 

HK acknowledges support by NASA under grant NAG5-12974.

\clearpage

\clearpage

\clearpage

\clearpage

\clearpage

\clearpage

\clearpage

\clearpage

\clearpage

\clearpage

\clearpage




\clearpage
\begin{table}
  \centering
  \begin{tabular}{c c}
    Parameter & Value \\
    \hline\hline\\
    A & -7.05\\
    B & 1.30\\
    C & -0.034\\
    $D_{1}$ & 0.057\\
    $\alpha$ & -0.20\\
    $D_{2}$ & -1.96\\
    $\beta$ & 2.44\\
    $d_{0}$ & 0.75$^\circ$\\
    \hline
  \end{tabular}
  \caption{\label{parameters}Parameters used for the TeV \gray energy estimator.}
\end{table}

\clearpage

\begin{table}
\centering
\begin{tabular}{c c c c c c c c c c}
$\delta$ & B (G) & R (m) &$\mathrm{log \, \gamma_{min}}$ &$\mathrm{log \, \gamma_{max}}$ &$\mathrm{log \, \gamma_{b}}$ &$U_{B}$ &
$U_{part}$ & $\mathrm{U_{part} \over U_{B}}$ & $\mathrm{L_{k}}$ \\
\hline  \\
50 & 0.2 & $1.05 \times 10^{13}$ & 3.3 & 5.3 & 4.8 & $1.92 \times 10^{-3}$ & $8 \times 10^{-2}$ &
41.5 & 2.12 \\
1000 & 0.5 & $1.7 \times 10^{10}$ & 2.8 & 4.4 & 3.26 & $9.95 \times 10^{-3}$
& 400 & $4.02 \times10^{4}$ & 10.9 \\
\hline
\end{tabular}
\caption{\label{fits}Parameters for 2 synchrotron self-Compton models. 
  Here, $\delta$ is the relativistic Doppler factor, B is the magnetic
  field, in Gauss, R is the size of the emission region, in meters,
  $\gamma_{min}$,$\gamma_{max}$ and $\gamma_{b}$ are the starting, ending,
  and break Lorentz factors for the primary electron energy spectrum,
$\mathrm{U_{B}}$ and $\mathrm{U_{part}}$ are energy densities
  in units of $\mathrm{erg \, \, cm^{-3}}$, and L$_{k}$ is the minimum
  kinetic luminosity (defined in section 5) in units
  of $10^{43}$ erg s$^{-1}$.}
\end{table}

\end{document}